\documentstyle[psfig]{caosp}
\begin{document}
\pubyear{1998}
\volume{27}
\firstpage{238}

\htitle{ Central Asian Network ...}
\hauthor{D.E. Mkrtichian {\it et al.}}

\title{Central Asian Network (CAN) -- the history and present status}
\vskip 2cm
\author{D.E. Mkrtichian \inst{1} \and A.V. Kusakin \inst{2} \and
E.B. Janiashvili, \inst{3} \\ J.G. Lominadze \inst{3} \and K. Kuratov \inst{4}
V.G. Kornilov \inst{2} \and N.I. Dorokhov \inst{1} \and  S. Mukhamednazarov \inst{5}}

\institute{Astronomical Observatory, Odessa State University,
Shevchenko Park, Odessa, 270014, Ukraine \and
Sternberg State Astronomical Institute, Universitetsky prospect, 13
Moscow, 119899, Russia  \and  Abastumani Astrophys. Observatory,
Abastumani, Mt. Kanobili, Georgia \and Tien Shan Astron. Observatory
and Fesenkov Astrophys. Institute of the Kazakhstan
National Acad. of Sci., Kamenskoe plato,  Almaty, 480068, Kazakhstan
\and Institute of Physics and Technology of Turkmenian Academy of Sciences
and Mt. Dushak-Erekdag Observatory, Ashgabad, Turkmenia}
\maketitle
\section{ Creation of CAN}
The necessity of long-term and continuous data for precise
determination of frequency spectra of pulsating stars
activated the creation of the asteroseismological networks of
observatories with a good longitude coverage such as
WET, DSN, STEPHI, STACC etc.

However, the Central-Asian gap in distribution of Northern
observatories  working in the known asteroseismological networks did not
allow a safe continuous time coverage in the Central Asian region during
the campaigns, especially in summer time.

In 1994, as a result of
the  agreement concluded between the team of researchers
and administrations of four (Euro)Asian observatories: Astronomical
Observatory of Odessa State University (AAOSU, Ukraine) and Mount
Dushak-Erekdag Observatory (MDEO, Turkmenistan), Tien-Shan
Astronomical Observatory  (TSAO, Kazakhstan) and Abastumani
Astrophysical Observatory (AbAO), the  informal Central Asian Network (CAN) of
researchers and observatories was created.

\section{\bf The CAN aims: }
\vspace{-3mm}

The main aim of the network created was carrying out coordinated
photometric work and  surveys on  pulsating stars mainly on Delta Scuti,
Lambda Boo and roAp stars visible in the Northern Hemisphere.
Participation of all CAN sites in the intercontinental campaigns of
other networks and groups is supposed to enhance the probability of
continuous monitoring in Central Asia due to uniform longitude coverage
of CAN observatories over 4 hour angles. In addition, the
network was aimed at photometric support of the  spectroscopic
projects on large telescopes.

\vspace{-3mm}

\section{The CAN sites:}

\vspace{-3mm}

In Table 1 are presented some brief informations on CAN sites.
   \begin{table*}
      \caption{ The CAN sites and photometric telescopes}
         \begin{center}
\footnotesize
\begin{tabular}{lccll}
\hline Observatory&Longitude    &Latitude &Telescope(s)&Photometers\\
\hline
MDEO       &$+57^\circ$51' &$+37^\circ$ 56' &1.0\,m  & 1024x1024 CCD camera *\\
           &          &  &0.8       &U,B,V,R,{\it v} (two-star)\\
TSAO       &+$76^\circ$ 57'& $+43^\circ$ 11' & 1.0\,m      &W,B,V,R  \\
           &          &  &0.48\,m (West.)& W,B,V,R \\
           &          &  &0.48\,m (East.)&W,B,V,R \\
AbAO       &$+42^\circ$ 49.5'&$+41^\circ$ 45.31'&1.25\,m&{\it u,v,b,y
$\beta$}\\
           &          &  &0.48\,m     &U,B,V,R \\
           &          &  &0.75\,m     & CCD camera \\
AOOSU      &$+30^\circ$ 45.5'&$+46^\circ$ 28.6'  &0.5\,m           &U,B,V,R \\
           &          &  &0.48\,m          &CCD camera, U,B,V,I,J,K\\
\hline
\end{tabular}
\end{center}
\vspace{-3mm}
{\small * CCD camera will be used from 1998 on, within the framework
of a Vienna(IfA) - Odessa (AOOSU) cooperation.}
 \end{table*}

\vspace{-3mm}

\subsection{Mount Dushak-Erekdag Observatory, Turkmenistan}

\vspace{-2mm}

The MDEO is jointly operated by the Institute of Physics and Technology of
the Turkmenian Academy of Sciences, Turkmenistan and the Astronomical
Observatory of the Odessa State University (Ukraine) (Dorokhov et al., 1995).
The observatory is
located at an altitude of 2020m on the slope of Mount Dushak-Erekdag, 5 km
south-west and 450 m below the summit. Geographically, Mount Dushak
Erekdag represents a single summit located directly at the border between
the Kopet-Dag Mountains and the Kara-Kum Desert. The distribution of
clear nights reaches its maximum in the summer-autumn season. According to
the weather statistics between 1992 and 1997, the number of photometric
nights during this period was about 80-90 {\%}.
The telescopes available for photometry
are the 1.0 m Schwarzschild-type fast and wide-field telescope (1:1.8) of
the Turkmenian Academy of Sciences, the 0.8m Ritchey-Chretien telescope of
the Odessa Astronomical Observatory and the double-tube 2\,x\,0.5\,m
Cassegrain telescope of the Turkmenian Academy  of Sciences.

\vspace{-2mm}

\subsection{Tien-Shan Astronomical Observatory, Kazakhstan }

\vspace{-2mm}

The TSAO is situated in the middle of Central Asia, within the
Tien Shan mountain chain, 2800\,m above sea level.
The distribution of clear nights at TSAO reaches its maximum in autumn
and winter, when a clear sky is rare at most European observatories and
well complements the clear period in Europe. TSAO is jointly operated by
the Fesenkov Astrophysical Institute of the National Academy  of Sciences
of  Kazakhstan  and the Sternberg State Astronomical Institute (Russia).
The facilities available  for photometry are the 1.0\,m Zeiss
Ritchey-Chretien and two 0.48\,m  Cassegrain telescopes with  two
4 channel W,B,V,R photometers attached to them.

\vspace{-1mm}

\subsection{Abastumani Astrophysical Observatory, Georgia}

\vspace{-1mm}

The Abastumani Astrophysical Observatory (AbAO) of the Academy of
Sciences of Georgia  is placed in South-West Georgia between the ranges of
Small Caucasus, at an altitude of 1700\,m above sea level.  The distribution
of clear nights at AbAO amounts  to its maximum in summer-autumn and winter
periods.  The 1.25\,m Ritchey Chretien telescope equipped with a two-star
photometer, the 0.7\,m Maksutov and the 0.48\,m Cassegrain telescopes are
available for photometric observations.

\vspace{-1mm}
\subsection{Astronomical Observatory of Odessa State University, Ukraine}
\vspace{-1mm}
For extending the CAN sites to the West during  the CAN
campaigns, we used the additional Western tracking site in Ukraine - the
Mayaki Station of the Astronomical Observatory of
the Odessa State University (AOOSU).
The Mayaki Station is located 45 km north-west of Odessa.
The climate is typical for Eastern Europe with a maximum of
clear nights in the August-September period. Two photometric  0.48m and
0.5m telescopes are available for photometry.

\vspace{-1mm}

\section{Activity of CAN}

\vspace{-2mm}

   Since the date of creation of CAN and during its 1994-1996 activity,
mainly the Eastern wing of the network (MDEO -- TSAO) was working,
while from 1997 on, the Eastern and Western (MDEO -- AbAO) wings
 began working jointly. According to the strategy of CAN we organized and
took part in the campaigns of other groups on the following  projects for
different types of pulsating stars:

\vspace{-2mm}
\subsection{Ap stars:}
\begin{itemize}
\item  {\bf 1994, September:}  The  Northern sky rapid photometry survey of
faint Ap stars in Kapteyn Selected Areas (PI, D. Mkrtichian) was started at
MDEO and continued in 1995-1996 at TSHAO (PI, A. Kusakin).
\item {\bf 1994 September:} A two-site photometric (MDEO) and radial velocity
(Simeiz Observatory, Ukraine) mini-campaign on the roAp star $\gamma$\,Equ (PI,
D.E. Mkrtichian) was  carried out.
\item {\bf 1994, September:}  MDEO and TSAO took part in a multisite
photometric campaign on the chemically peculiar star ET And
(PI, W.W. Weiss, IfA, Vienna)
\item {\bf 1995, March:} The CAN site TSHAO (observer, A. Kusakin) and
Konkoly Observatory (observer, M. Papar\'o) supported the spectroscopic work
made with the SAO 6.0\,m telescope on the chemically peculiar SrCrEu star
49 Cam (PI, D.E. Mkrtichian, AOOSU).
\end{itemize}

\subsection{$\delta$\,Scuti stars :}
\begin{itemize}
\item {\bf 1994, November:}   MDEO and TSAO took part in the DSN\,12 campaign
on the $\delta$ Scuti-type star $\theta^2$\,Tau (PI, M.Breger, IfA, Vienna).
\item {\bf 1994, November  and 1995, March: }  The CAN site TSHAO (obs.
A. Kusakin) and Konkoly Obs. (obs. M. Papar\'o) supported spectroscopic work
made with the SAO 6.0\,m telescope on the $\Delta$ Sct star VW Ari
(PI, D.E. Mkrtichian, AOOSU).
\item {\bf 1997, March - May:}  All CAN sites -- AbAO, MDEO and TSAO --
together with the Austrian 0.75 APT (Arizona) took part in
a 3-month (1 March--1 June) DSN\,17 multisite photometric campaign on
 the $\delta$ Scuti star 4 CVn
(PI, M. Breger,  IfA, Vienna). Data of 43 nights were acquired from CAN
sites in addition to 50 APT nights.
\item  {\bf 1997, September -- October:}  The multisite monitoring of
the eclipsing binary $\delta$ Scuti star AB Cas was carried out at AOOSU,
TSAO and AbAO.
\end{itemize}
\subsection{$\lambda$ Boo stars:}
\begin{itemize}
\item {\bf 1995/1996 :} The  survey of $\Lambda$ Boo stars was carried out at
TSHAO (Obs.: A. Kusakin) and 3 objects were observed to search for
pulsational variability.
\item {\bf 1995, July-September:}  Photometric work was devoted to the
$\Lambda$ Boo star 29 Cyg by using TSAO as the basic observing site
(Kusakin \& Mkrtichian, 1996).
\item {\bf 1996, July-September:}  The first, July-September multisite
spectroscopic
and photometric campaign on 29 Cyg was undertaken by CAN (PI, D. Mkrtichian)
the main observing sites being TSAO, Crimean Astrophysical Observatory
(CrAO), Konkoly Observatory, Ege University Observatory,
University of Toronto Observatory and Hawaii University Observatory
(Mkrtichian et al., 1998). 48 photometric and 2 spectroscopic nights
 were acquired.
\item {\bf 1997, July - October} - The second  multisite photometric and 
spectroscopic campaign on the pulsating $\lambda$\,Boo star 29\,Cyg
coordinated by CAN (PI, D. Mkrtichian) was undertaken with
the participating observatories AOOSU, TSAO, CrAO,  TUBITAK Observatory
(Turkey), AbAO and Austrian APT (Arizona).
 More than 80 photometric and 2 simultaneous spectroscopic nights were
 acquired.
\end{itemize}


\section{Further  work}

The three years of CAN activity have shown the expected efficiency of the 
network, satisfactory weather conditions and the good quality of photometry
at each site. Further strategy of CAN was re-discussed by PI's in early 1997
and aimed at:
\begin{itemize}
\item  Continuation of deep asteroseismological investigation and data
collection on selected Lambda Boo stars.
\item  Continuation of the survey of roAp stars in the Northern sky.
\item  Search for pulsating components in eclipsing binary systems
 and investigations of them.
\item  Participation in intercontinental multisite campaigns of other
groups.
\end{itemize}
  According to the newly developed CAN strategy, we have the following
schedule of CAN projects in the immediate future:
\begin{itemize}
\item {\bf 1998:} - Third multi-site photometric and spectroscopic campaign
on the pulsating $\lambda$\,Boo star 29 Cyg (PI, D. Mkrtichian)
\item {\bf 1997/1998:} - Monitoring of the $\delta$ Scuti stars AB Cas, Y Cam 
and V577 Oph, which are members of known eclipsing systems (PIs, D. Mkrtichian,
A. Kusakin).
\item {\bf 1997/1998:} Search for new pulsating A-F components
of selected eclipsing binary systems (PI, E. Janiashvili).
\item {\bf 1998:} Continuation of the survey of roAp stars in
the Northern sky, in Selected Kapteyn Areas and in associations
(PI, D. Mkrtichian).
\end{itemize}


\section{Collaboration with other groups}
CAN is open to suggestions from other groups for participation in their
campaigns. Requests for detailed information on the activity of CAN,
forthcoming campaigns and suggestions can be addressed to MDE at the e-mail
address: david@oao.odessa.ua.


\end{document}